# Non-Hermitian chiral degeneracy of gated graphene metasurfaces

Soojeong Baek[1,†], Sang Hyun Park[2,†], Donghak Oh[1,†], Kanghee Lee[1], Sang-Ha Lee[1], Hosub Lim[3], Taewoo Ha[4], Hyun-Sung Park[5], Shuang Zhang[6], Lan Yang[7], Bumki Min[1,8,*], Teun-Teun Kim[9,*]

[1]*Department of Mechanical Engineering, Korea Advanced Institute of Science and Technology (KAIST), Daejeon 34141, Republic of Korea*

[2]*Department of Electrical and Computer Engineering, University of Minnesota, Minneapolis, MN 55455, USA*

[3]*Harvard Institute of Medicine, Harvard Medical School, Harvard University, Brigham and Women's Hospital, Boston, Massachusetts 02215, United States*

[4]*Center for Integrated Nanostructure Physics, Institute for Basic Science (IBS), Sungkyunkwan University, Suwon 16419, Republic of Korea*

[5]*Samsung Advanced Institute of Technology, Samsung Electronics, Suwon 16678, Republic of Korea*

[6]*Department of Physics, University of Hong Kong, Hong Kong, 999077, China*

[7]*Department of Electrical and Systems Engineering, Washington University, Saint Louis, MO 63130, USA*

[8]*Department of Physics, Korea Advanced Institute of Science and Technology (KAIST), Daejeon 34141, Republic of Korea*

[9]*Department of Physics, University of Ulsan, Ulsan 44610, Republic of Korea*

[*] Correspondence to: bmin@kaist.ac.kr or ttkim@ulsan.ac.kr.

[†]These authors contributed equally to this work.

## Abstract

Non-Hermitian degeneracies, also known as exceptional points (EPs), have been the focus of much attention due to their singular eigenvalue surface structure. Nevertheless, as pertaining to a non-Hermitian metasurface platform, the reduction of an eigenspace dimensionality at the EP has been investigated mostly in a passive repetitive manner. Here, we propose an electrical and spectral way of resolving chiral EPs and clarifying the consequences of chiral mode collapsing of a non-Hermitian gated graphene metasurface. More specifically, the measured non-Hermitian Jones matrix in parameter space enables the quantification of nonorthogonality of polarisation eigenstates and half-integer topological charges associated with a chiral EP. Interestingly, the output polarisation state can be made orthogonal to the coalesced polarisation eigenstate of the metasurface, revealing the missing dimension at the chiral EP. In addition, the maximal nonorthogonality at the chiral EP leads to a blocking of one of the cross-polarised transmission pathways and, consequently, the observation of enhanced asymmetric polarisation conversion. We anticipate that electrically controllable non-Hermitian metasurface platforms can serve as an interesting framework for the investigation of rich non-Hermitian polarisation dynamics around chiral EPs.

A physical system describable with a non-Hermitian Hamiltonian may host exceptional points[1–4] (EPs), i.e., branching point singularities at which two or more eigenstates coalesce in parameter space. Unlike the degeneracies in Hermitian systems, for which an orthogonal set of eigenstates can be constructed, the eigenstates coalesce at the EP and become self-orthogonal, leading to a defective eigenspace of reduced dimensionality. These singular features have been observed and utilised in various quantum and classic systems, including electronic spins[5], superconducting qubits[6], condensed exciton-polaritons[7], electronic circuits[8], thermotic systems[9] and active matter[10]. Particularly in photonic systems[11–14], the ease of precise loss and/or gain control has facilitated the discovery of a plethora of EP-associated exotic behaviours, with some representative examples including chiral mode transfer with or without encircling around EPs[15,16], controlled electromagnetically induced transparency[17], a ring (or a pair) of EPs in momentum space[18,19], and coupling to the missing dimension at an EP[20]. In line with these advancements, we have also witnessed a series of promising EP-enabled functionalities, such as parity-time (PT) symmetry-broken lasing[21], exceptional topological phase engineering[22], electrical winding number switching[23], exceptional sensing[24,25] and coherent perfect absorption[26].

For the exploration of non-Hermitian physics and the application of EP-enabled functionalities, metasurfaces[22,23,27–29] are now being considered one of the most versatile platforms because their constituent meta-atoms are inherently constructed from lossy coupled subwavelength-scale resonators. Generally, any change in the polarisation state of light transmitted through the non-Hermitian metasurface can be characterised by a non-Hermitian Jones matrix that plays the role of an effective Hamiltonian[28–33]. In contrast to the prevailing cases[34], the non-Hermitian Jones matrix, of which the complex-valued elements can be engineered by geometrical and materials design of the meta-atoms, enables the utilisation of polarisation eigenstates for the examination of EP-related phenomena. Interestingly, at THz frequencies, the metasurface platform has been the only one that allows for the experimental observation of EPs, inheriting all the generic advantages of subwavelength-scale metaphotonics. However, until now, experimental probing of a branching point singularity in the parameter space has mostly been demonstrated in a passive way by repeatedly fabricating metasurfaces with varying meta-atom designs[28,29]. Furthermore, even with a series of repeated preparations, unavoidable errors from fabrications and/or measurements have made it difficult to observe relevant non-Hermitian dynamics around/at EPs. It is thus highly desirable to have precise real-time control of the parameters for access to an EP in a single metasurface platform[23,35].

To circumvent the aforementioned problems, we hybridise gated graphene microribbons with non-Hermitian metasurfaces and demonstrate the electrically controlled probing of polarisation eigentransmission surfaces along with the corresponding eigenstates. Notably, this probing methodology utilises time domain spectroscopy that makes use of a broadband pulse, which in combination with a continuous gate tuning capability enables high-resolution access to chiral EPs in two-parameter space. Here, chiral EPs refer specifically to the non-Hermitian degeneracy at which a circularly polarised state becomes the only eigenstate as a result of coalescence. The measured non-Hermitian Jones matrix in the parameter space enables a systematic investigation of nonorthogonality between polarisation eigenstates and atypical linkage between input and output polarisation states at the chiral EP. Specifically, we show that, for a specific incident polarisation, augmenting dimensionality at the chiral EP can be solely revealed at the output. It is also found that the maximal nonorthogonality assured by the defective Jones matrix at the chiral EP leads to the observation of enhanced asymmetric polarisation conversion. Last but not least, the examination of polarisation eigenstates in parameter space reveals a vortex structure, from which half integer topological charges at the chiral EP are clarified.

## Results

**Design of non-Hermitian gated graphene metasurfaces**. To map eigentransmission surfaces and investigate their structure near a chiral EP, we designed a non-Hermitian gated graphene metasurface consisting of an array of pairs of coupled split ring resonators (SRRs) with a graphene microribbon bridging the SRRs (Fig. 1a). The paired SRRs have their splits opened in orthogonal directions and are characterised by distinct external loss rates (Fig. 1b). Then, by employing temporal coupled-mode theory (TCMT)[31,33,36], a parameter-dependent non-Hermitian Jones matrix of the designed metasurface can be derived (see Methods). The two coupled SRRs are modelled as two orthogonally oriented resonators with a resonance frequency of $\omega_0$, a coupling rate of $\kappa$, and intrinsic and external loss rates of $\gamma_{i\mu}$ and $\gamma_{e\mu}$ ($\mu = x, y$), all of which can be adjusted to a certain degree by the geometry and materials constituting the unit cell (Fig. 1a, c). Under steady-state conditions, a 2×2 non-Hermitian Jones matrix $\mathbf{T}_l$ can be written in a *linear* polarisation basis. The matrix can be expressed as a sum of uncoupled and coupled parts ($\mathbf{T}_l = \mathbf{T}_{lu} + \mathbf{T}_{lc}$), only the latter of which is relevant for investigating the coalescing behaviour near/at the chiral EP. Specifically, the coupled part is found to be proportional to the following matrix:

$$\mathbf{T}_{lc} \propto \begin{bmatrix} \Omega_y + j\Gamma & K \\ K & \Omega_x - j\Gamma \end{bmatrix} \qquad (1)$$

where the dimensionless parameters are introduced for the simplicity of expression (see Methods for details): $\Omega_y = (\omega_0 - \omega)/\gamma_{ey}$, $\Omega_x = (\omega_0 - \omega)/\gamma_{ex}$, $\Gamma = (\gamma_{iy}/\gamma_{ey} - \gamma_{ix}/\gamma_{ex})/2$, and $K = -\kappa/\sqrt{\gamma_{ex}\gamma_{ey}}$. An inspection of the eigenvalues and eigenvectors of the above matrix shows the presence of a pair of chiral EPs when the following conditions are satisfied:

$$\Omega_x = \Omega_y, \qquad \Gamma = \pm K. \tag{2}$$

The first equality specifies a one-dimensional subspace satisfying PT symmetry, while the second equalities further identify the two chiral EPs as singularities distinguishing a PT exact phase from a broken one on the subspace. The chirality of a coalesced eigenstate at each EP is determined by the sign in the second equality (i.e., + for RCP and − for LCP). In this work, the system is parameterised by two variables: the angular frequency $\omega$ of an input wave and the gate voltage $V_g$ (or the Fermi level $E_F$ in simulations) that control the optical conductivity of graphene microribbons (Fig. 1c). The gating of graphene disproportionately adjusts the intrinsic loss rate of each SRR, resulting in a change in the $\Gamma$ value and correspondingly the potential fulfilment of the second equality (here, in this experimental work, $\Gamma = -K$). From the simulations, it is found that the external loss rates are almost invariant within the gating range of interest. It is worthwhile to note that the pair of chiral EPs connected by a PT broken phase in the parameter space of our metasurface platform is analogous to the pair of EPs linked by a bulk Fermi arc in the momentum space of a two-dimensional non-Hermitian photonic crystal[19,37,38].

**Mapping of the eigentransmission surface and identification of chiral EP** Non-Hermitian graphene metasurfaces were prepared by standard microfabrication techniques and a CVD-grown graphene transfer method and characterised by terahertz time domain spectroscopy (THz-TDS, see Methods). Here, a charge-neutral point of the graphene is located at a gate voltage of $\sim -1.1$ V. From the measured co- and cross-polarised complex amplitude transmission through the fabricated metasurface, a set of non-Hermitian Jones matrices $\mathbf{T}_l$, each of which is specified on a rectangular grid in the two-parameter space, can be obtained. The eigentransmission of $\mathbf{T}_l$ clearly reveals self-intersecting Riemann surface structures (Fig. 2a, b); the cusp at the end of the line of intersection is identified as the chiral EP and found to be located at $(\omega_{EP}, V_{g,EP})$ = (0.55 THz, 0.1 V). To support the experimentally measured topological structure, we performed numerical simulations using a finite element method and extracted the eigentransmission (see Methods). As shown in Fig. 2c, d, the numerical simulations show qualitative agreement with the experimental results. Interestingly, sectioning of the reconstructed eigentransmission surfaces further captures the key features of coupling and phase transitions across the chiral EP. First, a transition from weak ($\Gamma > -K$, $V_g > V_{g,EP}$) to strong ($\Gamma < -K$, $V_g < V_{g,EP}$) coupling between polarisation eigenstates can be seen by sampling eigentransmission surfaces at consecutively decreasing values of $V_g$ across the chiral EP (corresponding to three cut lines on the Riemann surfaces shown in Fig. 2a). As more clearly seen in Fig. 2e, f, a crossing (anti-crossing) to anti-crossing (crossing) transition is clearly observable in the spectrally resolved eigentransmission amplitudes (phases). Second, an exceptional phase transition is observed in the gate-voltage-dependent eigentransmission amplitudes (phases) sampled along a one-dimensional subspace satisfying PT symmetry ($\Omega_x = \Omega_y$); in these plots, exact and broken PT phases appear on either side of the chiral EP ($\Gamma = -K$, Fig. 2a-d). Here, it is also worth noting that the coupling crossover is observed to be concomitant with the exceptional phase transition across the chiral EP, as in other non-Hermitian systems[23,39].

**Nonorthogonality of eigenstates and half integer topological charge of chiral EP** To visualise the eigenstate coalescing behaviour, numerically calculated and experimentally extracted polarisation eigenstates are mapped on a Poincaré sphere (see Fig. 3a and Methods).

For clarity, the polarisation eigenstates corresponding to different values of gate voltage (or Fermi levels) are colour-coded. As seen in Fig. 3a, except at the chiral EP, the polarisation eigenstates exist in pairs and appear symmetrically with respect to the south pole. It is noteworthy that the paired polarisation eigenstates are not represented on the Poincaré sphere by antipodal points, which is indicative of the characteristic nonorthogonality of general non-Hermitian systems. As the chiral EP is approached in parameter space, the paired polarisation eigenstates move towards the south pole and eventually coalesce into the left circularly polarised state. To quantify the degree of nonorthogonality and coalescence, a Petermann factor ($F_p$) is calculated based on the left and right polarisation eigenstates extracted from the measurement (see Methods)[40,41]. Ideally, self-orthogonality and maximal nonorthogonality at the EP lead to the divergence of $F_p$, of which the experimental quantification can be done by plotting an inverse of $F_p$ in parameter space (Fig. 3b). In the plot, smaller values of $F_p^{-1}$ are seen along a one-dimensional subspace satisfying PT symmetry, and the value sharply drops down to ~$3\times10^{-4}$ near the chiral EP. This sharp decrease illustrates a singular sensitivity near the chiral EP to a variation in parameters. It is interesting to note that in addition to the investigation of nonorthogonality, the polarisation eigenstate mapping in parameter space enables the characterisation of topological charges associated with the chiral EPs. For this purpose, we monitored the cyclic variation of the ellipse orientation of polarisation eigenstates along an encircling path on the Riemann surface around the chiral EP[42–44] (Fig. 3c). The cyclic variation of the ellipse orientation (Fig. 3d) reveals a polarisation vortex centre at the chiral EP along with a half integer topological charge ($q = +1/2$). While not observable in the measurements due to the maximum gate voltage limit, the existence of the other chiral EP in parameter space with a half integer topological charge of $q = -1/2$ can be confirmed in the simulations[37,38,45].

**Revealing the missing dimension in a reduced polarisation eigenspace at chiral EP**

Polarisation eigenstate coalescence and the corresponding reduction of an eigenspace dimensionality at the chiral EP also lead to singular behaviours in the transmission of waves through the non-Hermitian metasurface. More specifically, the left and right polarisation eigenstates become self-orthogonal at the chiral EP so that the output polarisation state $|\psi_o\rangle$ is described with the single eigenpolarisation state $|L\rangle$ and the associated Jordan vector, i.e., in our case, $|J\rangle = |R\rangle$,

$$|\psi_o\rangle = (t_{LL,EP}|L\rangle\langle L| + t_{LL,EP}|J\rangle\langle J| + t_{LR,EP}|L\rangle\langle J|)|\psi_i\rangle \qquad (3)$$

where $t'$s are elements of the 2×2 non-Hermitian Jones matrix at the chiral EP written in a *circular* polarisation basis and $|\psi_o\rangle$ is the input polarisation state. Note that the matrix is in Jordan form at the chiral EP with its elements indicating cross- and co-polarised transmission ($t_{RR,EP} = t_{LL,EP}$ and $t_{RL,EP} = 0$). In Fig. 4, three illustrative cases are schematically shown along with their corresponding Poincaré sphere representations: (i) For RCP incidence ($|\psi_i\rangle = |J\rangle$, orthogonal to the polarisation eigenstate at the chiral EP, see Fig. 4a), the output polarisation state becomes a superposition of the polarisation eigenstate and the Jordan vector ($|\psi_o\rangle = t_{LL,EP}|J\rangle + t_{LR,EP}|L\rangle$). (ii) For LCP incidence ($|\psi_i\rangle = |L\rangle$, the polarisation eigenstate at the chiral EP, see Fig. 4b), the output polarisation state contains only the LCP component ($|\psi_o\rangle = t_{LR,EP}|L\rangle$). Note that the coalescence of polarisation eigenstates prohibits simultaneous nulling of both cross-polarised transmissions, which eventually leads to asymmetric polarisation conversion, as will be discussed below. (iii) Of particular interest is the case where the output polarisation state is completely devoid of the component parallel to the coalesced polarisation eigenstate (Fig. 4c); more specifically, preferential conversion to the Jordan vector ($|\psi_o\rangle = t_{LL,EP}|J\rangle$) can be achieved by setting the input polarisation states to $|\psi_i\rangle = -t_{LR,EP}/t_{LL,EP}|J\rangle +$

$|L\rangle$. This counterintuitive outcome is the accidental revelation of the missing dimension through the destructive interference of two LCP components: one from co-polarised transmission and the other from cross-polarised transmission of the prescribed input state. It is also worth mentioning that the solid angles subtended by output polarisation states are slightly smaller than those of input states (Fig. 4a-c) due to the nonunitary transformation performed by the metasurface[46]. More interestingly, the perfect nulling of cross-polarised transmission $t_{RL}$ at the chiral EP leads us to observe the signature of a Pancharatnam–Berry phase during gate-controlled coupling crossover[47] (Fig. 5a, b). The phase of $t_{RL}$ at high frequencies sharply changes by $2\pi$ (see Fig. 5b), implying zero-to-one topological winding number switching[23] by gating around the chiral EP (Fig. 5c). This winding number switching and the associated phase jump across the chiral EP can also be employed for enhanced sensing and monitoring of chemical and biological events[48].

**Maximal asymmetric polarisation conversion** The asymmetric non-Hermitian Jones matrix of the fabricated graphene metasurface also leads to gate-controlled asymmetric polarisation conversion (Fig. 5d). When compared with a forward propagation case, a backward propagating wave is incident on the metasurface consisting of unit cells that are mirror-reflected with a line of symmetry connecting the centres of two constituting SRRs. This guarantees that the off-diagonal elements of the Jones matrix are exchanged, i.e., $t_{LR}^b = t_{RL}^f$ and $t_{RL}^b = t_{LR}^f$, where superscripts specify the direction of propagation. Specifically, at the chiral EP of the fabricated metasurface, $t_{LR}^b$ becomes zero, while $t_{RL}^b$ remains nonzero, which suggests that a large difference in off-diagonal elements can be observed. To quantitatively characterise the effect, a normalised difference in asymmetric polarisation conversion is defined here as $\delta_t \equiv \left(\left|t_{RL}^b\right|^2 - \left|t_{RL}^f\right|^2\right) / \left(\left|t_{RL}^b\right|^2 + \left|t_{RL}^f\right|^2\right)$, the value of which generally ranges from -1 to 1. Figure 5e shows the values of parameter-dependent $\delta_t$ extracted from the transmission measurement performed on the fabricated metasurface. It is clearly seen that the values are nonzero near the chiral EP and approach the maximum value of 1 at the chiral EP. Therefore, the gate-controlled access to the chiral degeneracy enables us to reach the maximal asymmetric polarisation conversion that is hard to achieve in passive non-Hermitian metasurfaces due to the sensitivity of exceptional points to fabrication errors.

**Conclusions**

In this work, we experimentally demonstrated the potential of a non-Hermitian gated graphene metasurface platform for the clarification and characterisation of chiral EPs in parameter space. The proposed platform stands among other recently implemented tunable non-Hermitian photonic systems[23] while distinguishing itself from others by utilising a non-Hermitian Jones matrix for the manipulation of polarisation states. Specifically, in addition to the well-known general features such as nonorthogonality and mode coalescence, the non-Hermitian Jones matrix, especially written in Jordan form at the chiral EP, leads to an unusual nonunitary relation between input and output polarisation states. One such manifestation is the preferential polarisation conversion into the state represented by a Jordan vector of the non-Hermitian Jones matrix. This implies that the output polarisation state can be made independent of the coalesced eigenstate of the metasurface being transmitted, which is contrary to our usual conception. We further experimentally clarified half-integer topological charges of a non-Hermitian chiral degeneracy and topological winding number switching by gating. We believe that the proposed tunable metasurface platform may become an essential tool in the investigation of dynamic phenomena related to non-Hermitian chiral degeneracies and serve as a testbed for realising artificial non-Hermitian effective matter.

**Methods**

**Jones matrix of non-Hermitian metasurfaces** A parameter-dependent non-Hermitian Jones matrix can be obtained by modelling gated graphene metasurfaces with temporal coupled mode theory (TCMT). The analysis results in the following 2×2 non-Hermitian Jones matrix in a linear polarisation basis.

$$\mathbf{T} = \mathbf{I} - \mathbf{D}\mathbf{H}^{-1}\mathbf{D}$$

where $\mathbf{I}$ is the identity matrix, $\mathbf{D} = \begin{bmatrix} j\sqrt{\gamma_{ex}} & 0 \\ 0 & j\sqrt{\gamma_{ey}} \end{bmatrix}$, $\mathbf{H} = j(\mathbf{\Omega} - \omega\mathbf{I}) - \mathbf{\Gamma}_e - \mathbf{\Gamma}_i$, $\mathbf{\Gamma}_e = \begin{bmatrix} \gamma_{ex} & 0 \\ 0 & \gamma_{ey} \end{bmatrix}$, $\mathbf{\Gamma}_i = \begin{bmatrix} \gamma_{ix} & 0 \\ 0 & \gamma_{iy} \end{bmatrix}$, and $\mathbf{\Omega} = \begin{bmatrix} \omega_0 & \kappa \\ \kappa & \omega_0 \end{bmatrix}$. Then, the non-Hermitian Jones matrix can be expressed as

$$\mathbf{T} = \mathbf{T}_u + \mathbf{T}_c = \xi\mathbf{I} + \eta \begin{bmatrix} \Omega_y + j\Gamma & K \\ K & \Omega_x - j\Gamma \end{bmatrix}.$$

Here, for simplicity, we introduce two dimensionless parameters $\xi = 1 - \chi/\det(\mathbf{H})$ and $\eta = j\gamma_{ex}\gamma_{ey}/\det(\mathbf{H})$, where $\chi = \gamma_{ex}\gamma_{ey} + (\gamma_{ex}\gamma_{iy} + \gamma_{ey}\gamma_{ix})/2$. In a circular polarisation basis, the matrix can be written as

$$\mathbf{T}_{cir} = \mathbf{T}_{cir,u} + \mathbf{T}_{cir,c} = \xi\mathbf{I} + \frac{1}{2}\eta \begin{bmatrix} \Omega_x + \Omega_y & \Omega_x - \Omega_y + 2j(\Gamma - K) \\ \Omega_x - \Omega_y + 2j(\Gamma + K) & \Omega_x + \Omega_y \end{bmatrix}.$$

**Numerical simulations** To numerically calculate eigentransmission surfaces of the gated graphene metasurfaces, we use a commercial simulation tool that employs the commercial finite-element method solver CST Microwave Studio. In the frequency range of interest, the dielectric constant for gold is tabulated in ref.[49] and can be fitted by using the Drude model with a plasma frequency $\omega_p = 1.37 \times 10^{13}$ rad/s and a collision frequency $\gamma = 4.07 \times 10^{13}$ rad/s. The complex index of the silicon substrate is extracted experimentally by measuring the transmission of the THz wave through the substrate. The optical conductivity of graphene is modelled by Kubo's formula[50] with an experimentally fitted scattering time of 20 fs.

**Device fabrication** The gated graphene non-Hermitian metasurfaces are prepared by employing standard microelectromechanical fabrication techniques. All the metallic structures are made of 200-nm-thick gold and attached to the substrate with a 20-nm-thick chrome adhesion layer. To bridge the gap between SRRs with a graphene microribbon, CVD-grown graphene is first transferred to the substrate with previously patterned SRRs. The transfer of graphene is accomplished by using PMMA (C2, Microchem) as a supporting layer. The transferred large-area graphene is then patterned by UV lithography with bilayered photoresists (PMGI and HKT 501). After UV exposure and development, the part of graphene uncovered by photoresists is etched by a plasma asher. As shown in Fig. 1c, the graphene microribbon can be electrically doped by utilising an ion gel gate dielectric with in-plane gate and ground electrodes patterned on an undoped silicon substrate.

**THz-TDS measurement** To retrieve non-Hermitian Jones matrices, a conventional THz time-domain spectroscopy (THz-TDS) system is employed. The main part of the system consists of a Ti:sapphire femtosecond laser (Mai-Tai, Spectra-physics) operating at a repetition rate of 80 MHz with a centre wavelength of 800 nm, a GaAs photoconductive antenna (iPCA, BATOP) for the generation of a THz signal and a 1-mm-thick ZnTe crystal for detection. The generated

THz signal covers a spectral range from ~0.1 to ~2.0 THz. The co- and cross-polarised complex amplitude transmissions are measured by employing two wire-grid terahertz polarisers and used to retrieve the non-Hermitian Jones matrix.

**Acknowledgments**

This work was supported by National Research Foundation of Korea (NRF) through the government of Korea (NRF-2021R1C1C100631612 and 2017R1A2B3012364) and Institute of Information & communications Technology Planning and Evaluation (IITP) grant funded by the Korea government (MSIT) (No. 2022-0-00624). The work was also supported by the center for Advanced Meta-Materials (CAMM) funded by Korea Government (MSIP) as Global Frontier Project (NRF-2014M3A6B3063709). S.H.P acknowledge funding support from NSF/EFRI under Grant Agreement No. 1741660. The work is partially funded by NRF (NRF-2021R1A6A3A14044805, 2020R1C1C1009098) and the Institute for basic science (IBS-R011-D1).

**Author contributions**

S.B., S.H.P., S.Z., B.M., and T.-T. K conceived the original idea. S.B. developed the gated graphene metasurfaces and performed the THz-TDS measurement. S.B., S.H.P., and D.O. developed the analytical model. All authors analysed the data and discussed the results. S.B., S.H.P., D.O., S.Z., B.M., and T.-T.K. wrote the paper, and all authors provided feedback. B.M. and T.-T.K. supervised the project.

**Competing financial interests**

The authors declare no competing financial interests.

**Figure captions**

**Fig. 1 Design and fabrication of non-Hermitian gated graphene metasurfaces. a** Schematic illustration of the unit cell and its interaction with an incident wave. The unit cell is composed of two orthogonally oriented SRRs with an identical resonance frequency of $\omega_0$, distinct intrinsic and external loss rates of $\gamma_{i\mu}$ and $\gamma_{e\mu}$ ($\mu = x, y$), and a coupling strength $\kappa$. The two SRRs are bridged by a transferred single-layer graphene microribbon. **b** Simulated transmission amplitude through each SRR in the absence of the other SRR. The SRRs are excited by an incident wave polarised along the gap opening. The fitted external loss rates are $\gamma_{ex} = 0.53$ rad/s and $\gamma_{ey} = 0.03$ rad/s, respectively. **c** Schematic rendering and microscopic image of the non-Hermitian gated graphene metasurface. The geometric dimensions of the unit cell are set to $L_1 = 24\ \mu m, L_2 = 44\ \mu m, L_3 = 24\ \mu m, L_4 = 64\ \mu m, g_1 = 6\ \mu m, g_2 = 50\ \mu m, S = 2\ \mu m$ and $P_x = P_y = 120\ \mu m$. To apply gate voltage and change the conductivity of an array of graphene microribbons, a square-ring-shaped gate electrode, a ground electrode and an ion gel layer are incorporated into the non-Hermitian graphene metasurface. Here, two ends of the graphene microribbon are attached to the square-ring-shaped electrode, while the ion-gel layer covers both the gate and ground electrodes as well as the array of graphene microribbons.

**Fig. 2 Eigentransmission amplitude and phase of the non-Hermitian gated graphene metasurface. a** Experimentally measured eigentransmission amplitude

and **b** phase plotted in parameter space spanned by the input frequency and the gate voltage $V_g$ applied to the array of graphene microribbons. The chiral EP is denoted by a point (EP) on the surfaces. Chiral EP is observed at a frequency of 0.55 THz and a gate voltage of $V_{g,EP} = 0.1$ V. **c** Numerically calculated eigentransmission amplitude and **d** phase plotted in parameter space spanned by the input frequency and Fermi level $E_F$ of the graphene microribbons. The chiral EP is located at $\omega_{EP}$ = 0.54 THz and $E_F$ = 0.13 eV. **e** Eigentransmission amplitude plot sectioned at three values of gate voltage. A strong-to-weak coupling transition is clearly seen across the chiral EP. **f** Eigentransmission phase plot sectioned at three values of gate voltage.

**Fig. 3 Electrical access to eigenpolarisation states at the EP. a** Measured (square) and simulated (line) eigenpolarisation states plotted on the Poincaré sphere with LCP at the north pole with respect to $\omega$ and $V_g$. **b** Inverse of the Petermann factor ($F_p^{-1}$) calculated from eigenpolarisation states with respect to $\omega$ and $V_g$. **c** Cyclic parameter variation (black lines) around EP on Riemann surfaces of eigentransmission amplitude. Cynic-coloured and magenta-coloured dots indicate the starting points for the first and second round trip, respectively. **d** The variation of eigentrasnmission amplitude (top) and the corresponding eigenpolarisation ellipse angle (bottom) with respect to cyclic variation of $f$ and $V_g$ in parameter space.

**Fig. 4 Peculiar linkage between input and output states at the chiral EP**. **a-c** Conceptual schemes (left), input states (middle), and output states (right) on the Poincaré sphere for three polarisation input states at the chiral EP. **a** For the input state with RCP, the output polarisation states are superposed with RCP and LCP. **b** The

LCP component is transmitted only for the input state with LCP. **c** For specific input states where the LCP components destructively interfere with each other, the RCP component is only transmitted, and then the missing dimension is revealed.

**Fig. 5 Gate-controllable cross-polarised transmission *t*<sub>RL</sub> and asymmetric polarisation conversion. a-b** Transmission amplitude (**a**) and phase (**b**) for $t_{RL}$ in parameter space with incident frequency and gate voltage $V_g$. **c** Gate-tunable winding number, which indicates the number of times winds around the EP in the complex plane of $t_{RL}$. **d** Schematic view for asymmetric polarisation conversion of circularly polarised light through non-Hermitian graphene metasurfaces. **e** Normalised difference of intensity $\delta_t$ for the quantitative manifestation of asymmetric polarisation conversion with respect to $f$ and $V_g$.

# Figures

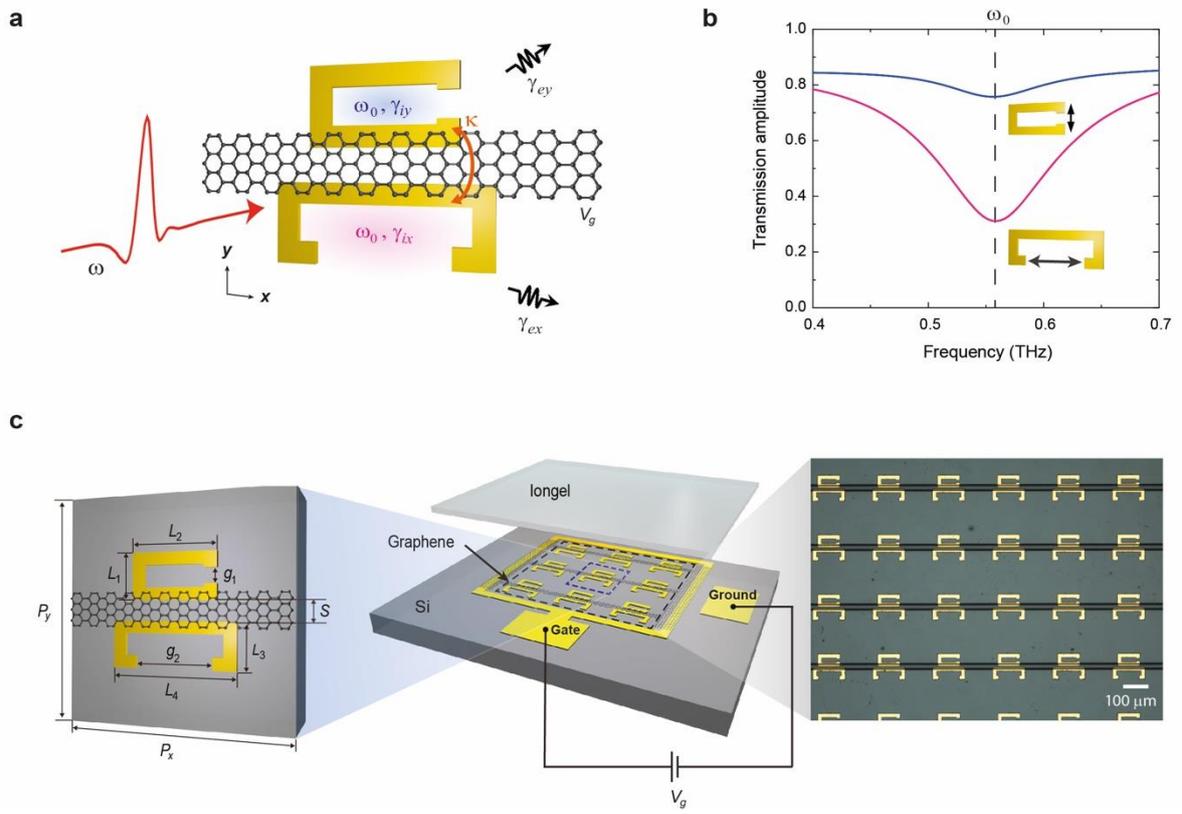

Fig. 1 Baek *et al.*

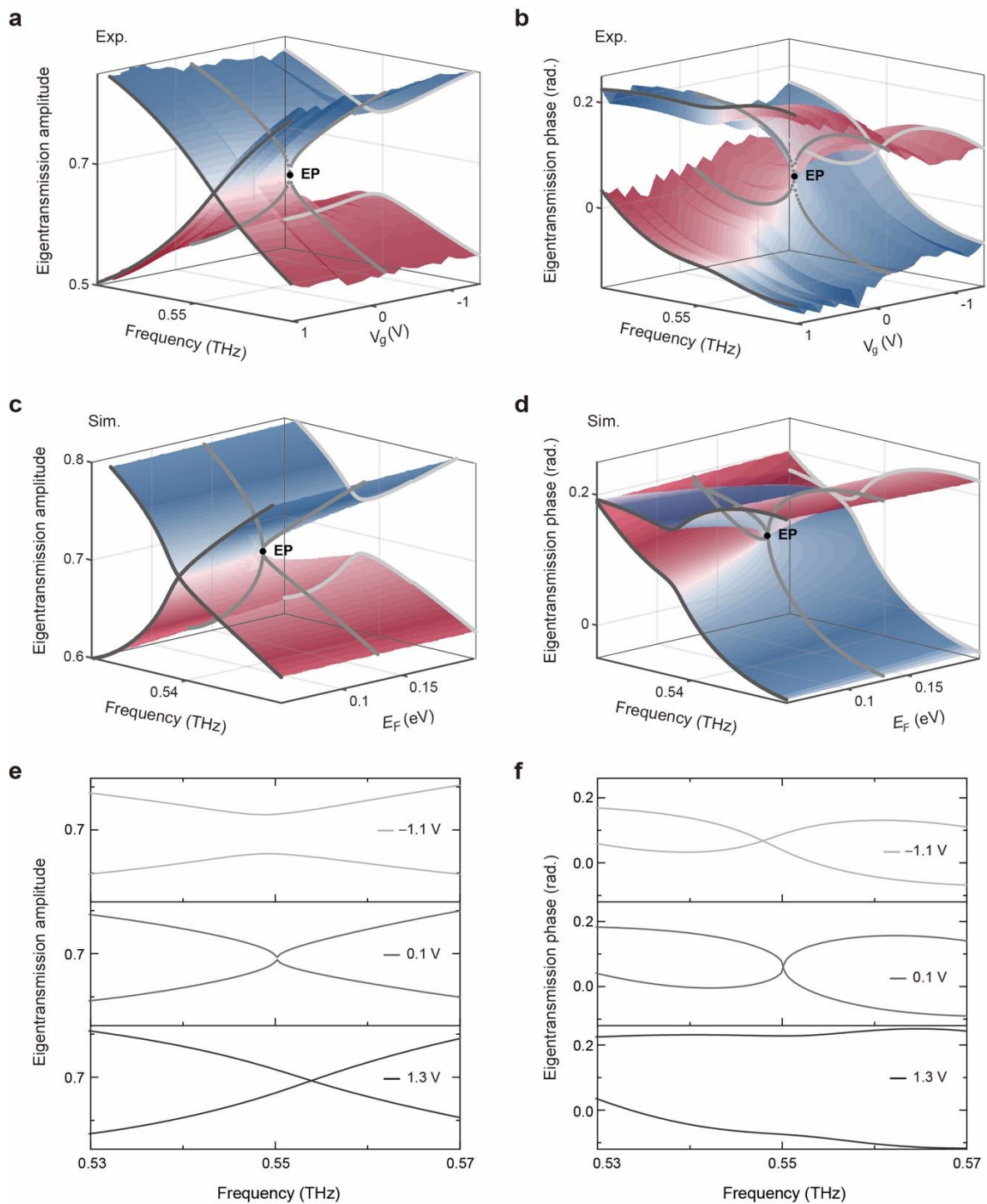

Fig. 2 Baek *et al.*

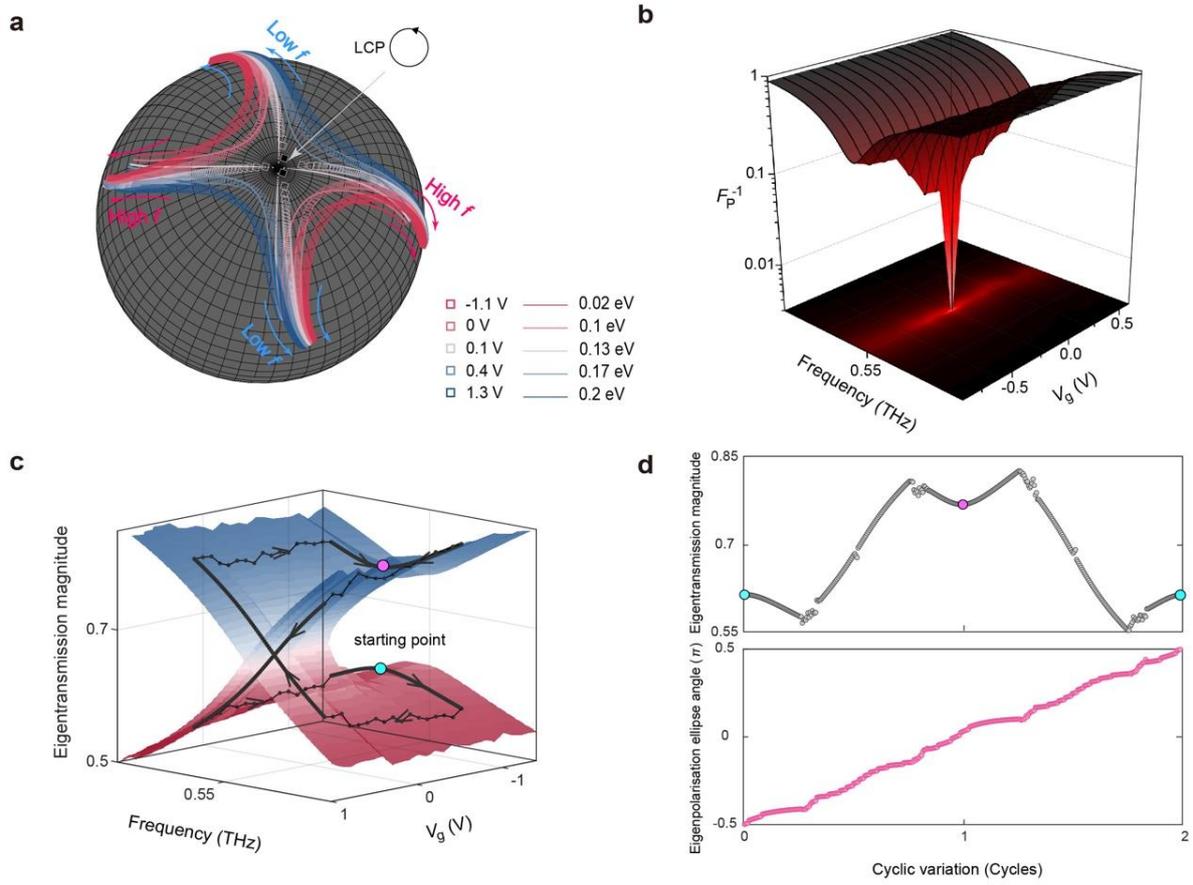

Baek *et al.* Fig. 3

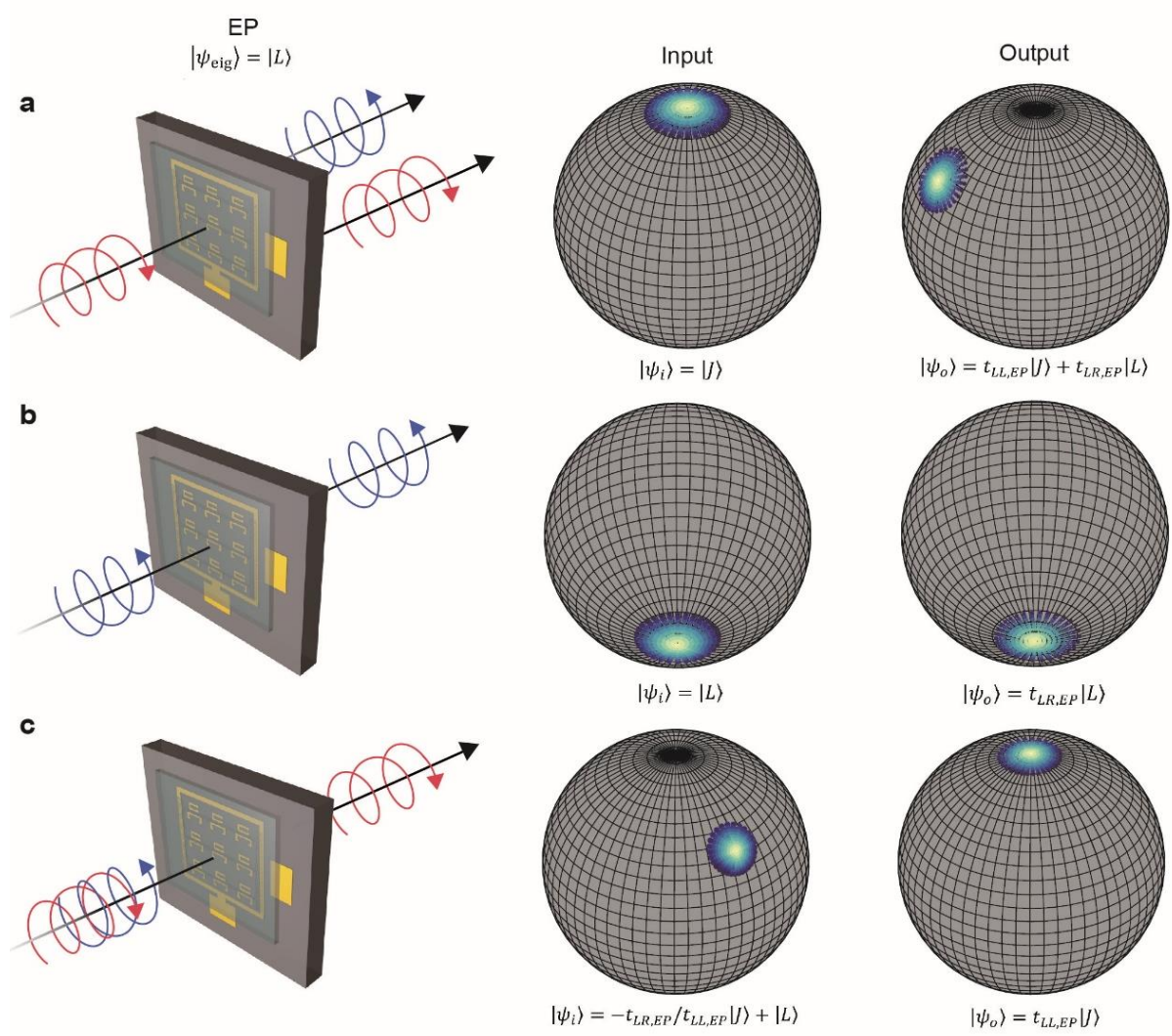

Baek *et al.* Fig. 4

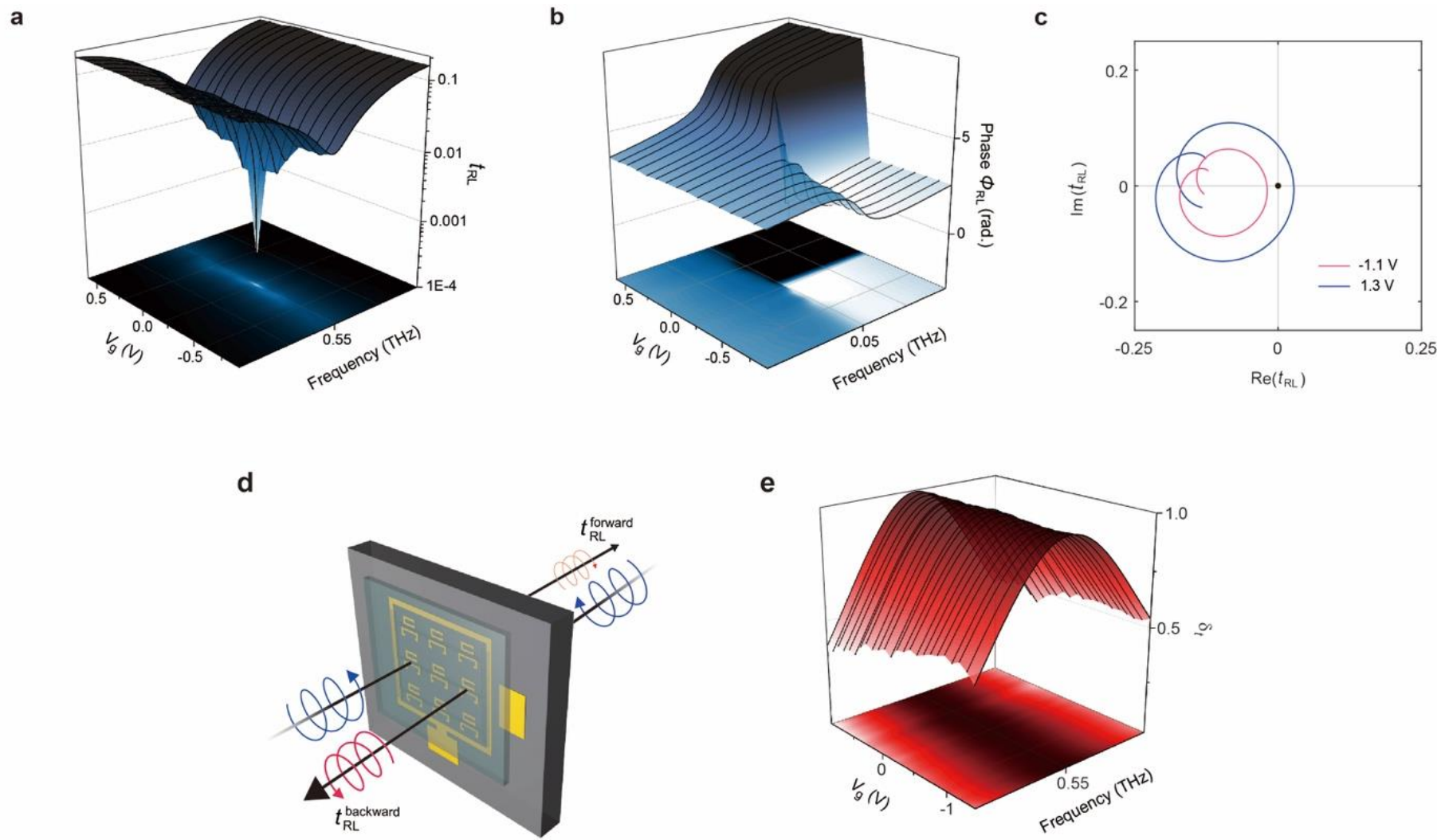

Baek *et al.* Fig. 5